\documentstyle[12pt,epsf]{article}
\begin{document}
%\draft
\begin{flushright}
Imperial/TP/96-97/29\\
{\tt hep-ph/9708??}\\
{\LaTeX}-ed on \today\\
\end{flushright}
%--------------------

\begin{center}
{{\Large \bf Initial Vortex Densities after a Temperature Quench}}
\end{center}
\vspace{.3in}

\begin{center}

{\large G. Karra$^\dagger$\footnote{email: g.karra@ic.ac.uk},
 R. J. Rivers$^\dagger$\footnote{email:r.rivers@ic.ac.uk}}
\bigskip

$^\dagger${\it Blackett Laboratory, Imperial College\\ London SW7 2BZ}\\

\bigskip

\end{center}
\vspace{.6in}

\thispagestyle{empty}

\begin{abstract}
We calculate the {\it initial} density of relativistic global
vortices (strings) produced 
at a quench and contrast it with the
predictions of Kibble and Zurek.
\end{abstract}

\vspace{0.2in}

\vskip2pc

Large-scale structure in
the universe has been attributed \cite{kibble1}
to the production of cosmic strings (field vortices) in the phase
transitions that are expected to occur in the Grand Unification era.  
However, it has been argued\cite{kibble2,zurek1,volovik} that
there are
similarities between the production of vortices in the early universe and in condensed matter
systems ($^4 He$ and $^3 He$ in particular), where there is more accessible experimental 
data\cite{lancaster,helsinki}.
It is this suggestion that we shall be addressing in this letter,
concentrating on one single property, the vortex density.

The simplest theory that displays vortices at a temperature quench is that
of a complex scalar field $\phi = (\phi_1 + i\phi_2)/\sqrt{2}$, 
determined by the $O(2)$-invariant action
\begin{equation}
S [\phi] = \int d^{4}x \biggl (
\frac{1}{2} (\partial_{\mu} \phi_a)^2  - \frac{1}{2}
m^{2}(t) \phi_a^2 - \frac{1}{4} \lambda (\phi_a^2)^2
\biggr ).
\label{St}
\end{equation}
We identify the effective mass
$m(t)$ in Eq.\ref{St} with the {\it
equilibrium} thermal mass, initially assumed positive. Once the
quench is over, $m^{2}(t)$ takes its zero-temperature physical value
$-M^{2}$.  In the interim, by reducing the temperature, we assume an explicit
time-dependence
$m^{2}(t) = -M^{2}t/t_Q$, if $-t_Q < t < t_Q$
and $m^{2}(t) =M^{2}$ when $t<-t_Q$ (and is equal to $-M^{2}$, for $t>t_Q$).
Adopting different initial and final mass values (i.e. different
initial temperatures $T_0 = O(M/\sqrt\lambda )$) will have little effect,
provided the initial mass is not too small.

Estimates of the initial vortex density can be made in several ways.
In Ref.\cite{kibble2} Kibble assumes that the {\it equilibrium}
correlation length $\xi_{eq}(t)= |m^{-1} (t)|$
describes
the correlation length $\xi(t)$ of the field until time
$t\approx -t_C$, where ${\dot\xi_{eq}}(-t_C ) = 1$, on causal grounds. If 
$\xi (t)$ is approximately frozen during the interval
$(-t_C ,t_C )$ then $t_C=t_{Q}^{1/3}(2M)^{-2/3}$,
at which time the number of {\it relativistic} strings per unit area is
\begin{equation}
n(t_C ) = O(1/\xi_{eq}^{2}(t_C )) = O\bigg((A t_{Q})^{-2/3}\bigg)
\label{tom}
\end{equation}
where $A = O(M^{-2})$ is the
cross-section of a {\it classical} vortex. 
The critical slowdown argument  presented by Zurek\cite{zurek1} sounds rather
different, but the results are qualitatively the same in their power
dependence on $t_Q$, supported by numerical simulations of the
time-dependent Landau-Ginzsburg equations \cite{zurek2}, the
standard dynamics of condensed matter systems.  This early
density is the largest as vortices subsequently annihilate each
other but the details of this are irrelevant to our problem. 
The  nonrelativistic counterpart
of Eq.\ref{tom} is in good agreement with the $^{3}He$ data\cite{helsinki}.
Distinct vortices require $n\ll A^{-1}$.

In this letter we examine whether the result (\ref{tom}),
based on the simple picture of \cite{kibble2}, or the
Landau-Ginzsburg equations of \cite{zurek2}, with its lack of direct
reference to quantum fluctuations, is general.
We shall see that the {\it weakly-coupled} quantum field theory of
Eq.\ref{St}, when
used directly to estimate vortex densities, gives an initial density
less than that of
Eq.\ref{tom}, and qualitatively different. 
%and Eq.\ref{nG}. 
The reason is that the time $t_C$ is
too early for a classical description of the field to be
appropriate.  This does not deny the usefulness of Eq.\ref{tom} for
condensed-matter systems. We first summarise our results and justify
them later. 

It is because the formation of defects is an early-time occurrence that
it is, in large part, amenable to analytic solution.  Nonetheless, the
production and freezing in of the vortices occurs over
several time scales, as is most easily seen for an {\it instantaneous}
quench ($t_Q =0$), for which Eq.\ref{tom} is clearly inappropriate,
a useful warning that the simple picture above has its limitations.   
However, once we have understood the instantaneous case the slow
quench is not much more difficult.

For the instantaneous quench there is, first, an interval
$Mt_r= O(1)$
before the field can respond.  This is followed by an interval
$Mt_{sp}= O[\ln (1/\lambda )]$
during which domains form
as a result of the growth of the long wavelength modes   as they fall from the top of the
potential hill towards its ground state value $\phi_{0}^{2} =
M^2 /\lambda $. 
After a further interval
backreaction (field self-interaction) will trap the field into one
of its vacuum values, 
labelled by the phase $\alpha$ of $\langle\phi\rangle$, at each
point in space. 
A consequence of this phase separation is that 
continuity and single valuedness will sometimes force the field to
remain in the false ground-state at $\phi = 0$, the core
of a tube of false vacuum, 
%of cross-section $A = O(M^{-2})$,
which characterises a vortex. By  identifying
string cores as lines of zeroes\cite{halperin} of the 
field $\phi$  at the spinodal time $t_{sp}$ we shall find
\begin{equation}
n(t_{sp}) = O\bigg(\frac{1}{A\ln (1/\lambda)}\bigg)
\label{ntsp}
\end{equation}
and argue that this is, qualitatively, the density after the backreaction has
taken effect.

As the quench time $t_Q$ is increased from zero we find different
behaviour according as $Mt_Q$ is $O(1)$, $O(\ln (1/\lambda ))$,
$O(1/\lambda)$ (or powers thereof), which we take
to be qualitatively distinct.
Most simply, $n$ of Eq.\ref{ntsp} is unchanged qualitatively as long as $Mt_Q = O(1)$.   However, once
$Mt_Q = 
O[(\ln (1/\lambda ))^{\alpha}]$
with $\alpha >1$, or even a small power of $1/\lambda$, the
situation begins to change.  The response time $t_r$ now satisfies
$Mt_r = O((Mt_Q)^{1/3})$, and
domains grow until a time $t_{sp}$, given by 
$Mt_{sp}= O[(Mt_Q )^{1/3}(\ln (1/\lambda ))^{2/3}]$.
The corresponding density is
\begin{equation}
n = O\bigg(\frac{1}{(At_{Q})^{2/3}(\ln (1/\lambda ))^{1/3}}\bigg) < n(t_{C}).
\label{enslow}
\end{equation}
Although this looks rather like $n(t_{C})$ of Eq.\ref{tom} appearances  are
deceptive, in that $\ln
(1/\lambda)$ is essentially a function of $Mt_{Q}$. 
%Finally, for quenches that are slower yet, the Ginzsburg regime
%begins to become important.  Even though the density of line
%zeroes $n_G$ of Eq.\ref{nG} is small, we doubt if they can be
%construed as vortices since their density depends strongly on the
%scale at which they are counted.  

We derive the results above 
by assuming Gaussian fluctuations. The vortex
density $n(t)$ at time $t$ after the quench is started is given in
terms of the equal-time Wightman function
with the given thermal boundary
conditions
\begin{equation}
W_{ab}(|{\bf x} -{\bf x} '|;t)  = \langle\phi_{a}(t ,{\bf
x})\phi_{b}(t ,{\bf
x}')\rangle  =\int d \! \! \! / ^3 k 
\, e^{i {\bf k} . {\bf x} } P(k;t)\delta_{ab}.
\end{equation}
$P(k;t)$ is the {\it power} of the field fluctuations with
wavevector ${\bf k}$.
Because of the non-equilibrium time evolution
there is no time translation and, assuming no initial bias,
$W_{ab}$ is $O(2)$ diagonal, $W_{ab}=\delta_{ab}W$.

We count
vortices by a judicious count of the line zeroes of the complex field. 
Care is needed since
quantum fluctuations lead to line zeroes
of the complex field on all distance scales. 
Let us cut off the field fluctuations at a wavenumber $\Lambda$.
In our Gaussian approximation, with $\langle\phi_{a}\rangle = 0$ at
all times, this is {\it equivalent} to integrating over
the 'environment' of short wavelength modes with $k>\Lambda$ to
construct a coarse-grained density matrix, from which $W_{\Lambda}$
is derived.
The resulting density of line zeroes
is\cite{halperin}
\begin{equation}
n_{\Lambda}(t)=\frac{-1}{2\pi}\frac{W''_{\Lambda}(r;t)}{W_{\Lambda}(0;t)}.
\label{nlambda}
\end{equation}
Of course, there will always be small-scale structure on vortices but
only when the density is insensitive to $\Lambda = O(M)$,
corresponding to their classical thickness, do we accept the
line zeroes as representing vortices with a classical future. 
That is, we require the power $k^2 P(k;t)$ 
in the fluctuations at $\Lambda = O(M)$ to be relatively small.

We return to the
{\it instantaneous} temperature quench ($t_Q = 0$) in which  the 
effective mass $m^{2}(t)$ flips from $+M^{2}$ to $-M^{2}$ at $t=0$.
Initially, let us calculate the Gaussian $W_{\Lambda}(r;t)$
for a {\it free roll}\cite{alray,guth},
built from the modes $\chi^{\pm}_{k}(t)$,  
\begin{equation}
\Biggl [ \frac{d^2}{dt^2} + {\bf k}^2 + m^2(t) \Biggr ]\chi^{\pm}_{k}(t)  =0,
\label{mode1}
\end{equation}
subject to $\chi^{\pm}_{k}(t)
=
e^{\pm i\omega_{in}t}$ at
$t\leq 0$,
for incident frequency $\omega_{in} = \sqrt{{\bf
k}^{2} + M^{2}}$. 
Imposing the KMS  boundary conditions at
$t\leq 0$ for an initial Boltzmann
distribution at temperature $T_0$
 determines  $W_{\Lambda}(r;t)$ uniquely. For $\Lambda=O(M)$, but $\Lambda >M$,
say, $W_{\Lambda}(r;t)$ can be decomposed in an obvious way as
$W_{\Lambda}(r;t) = W^{in}_{\Lambda}(r)$ for $t\leq 0$, and
\begin{equation}
W_{\Lambda}(r;t) = W^{in}_{\Lambda}(r) + W_{|{\bf k}|<M}(r;t) + W_{\Lambda >|{\bf k}|>M}(r;t)
\label{Wdecomp2}
\end{equation}
for $t>0$, where\cite{boyanovsky} the second and third terms are the long and
short wavelength fluctuations that grow from $t=0$ onwards.

Initially, fluctuations are so large that the density depends
critically on the scale $\Lambda^{-1}$ at which we view the field.  
The equilibrium background term $W_{\Lambda}^{in}(r)$ has the form
\begin{equation}
W_{\Lambda}^{in}(r)=\int_{|{\bf k}|<\Lambda } d \! \! \! / ^3 k 
\, e^{i {\bf k} . {\bf x} } C(k),
\end{equation}
where $ C(k) =\coth(\omega_{in} (k)/2T_0
)/2\omega_{in}(k)$ encodes the initial conditions.
The correlation length of the field is $M^{-1}$ and, for cutoff
$\Lambda = O(M)$, the cold vortex thickness, we have
$n_{M} =  O(M^2 )$.  There is a sea of interlocking line zeroes, 
separated only by the
Compton wavelength.  The oscillating field fluctuations make them
unsuitable  as serious candidates for string since,  
if we evaluate
$\partial
n_{\Lambda}/\partial\Lambda$ we find
\begin{equation}
\Theta_{M}(t)\equiv\frac{M}{n_{M}(t)}\frac{\partial
n_{\Lambda}(t)}{\partial\Lambda}\bigg|_{\Lambda = M} = O(1).
\end{equation}  
A density of line zeroes so dependent on the scale at which it 
is measured cannot be compared usefully with that of (\ref{tom}).
We term these the
{\it fluctuation vortices}. 

For early positive times $n_{\Lambda}$ is equally sensitive to
$\Lambda$ but, 
once $Mt\gg 1$, the relevant term is $W_{|{\bf k}|<M}(r;t)$,
whose integral at time t is dominated by the nonperturbatively large
peak\cite{boyanovsky} in the power of the fluctuations at $k$ around
$k_0$, where
$t k_0^2\simeq M$.  Once $k_{0}\ll M $ we have the required insensitivity.
We find, approximately, that
\begin{equation}
W_{|{\bf k}|<M}(r;t)
\propto\frac{MT_0}{(tM)^{3/2}}e^{2Mt}\exp\{-r^{2}M/4t\}.
\label{Wapp}
\end{equation}
The prefactor comes from approximating $C(k_{0})$ by  $T_0/M^{2}$.
For $t>0$ the
unstable modes grow until $W_{|{\bf k}|<M}(0;t_{sp}) =O(\phi^{2}_{0}) = O(M^{2}/\lambda )$, determining
the spinodal time $t_{sp}$  as   $Mt_{sp}=O(\ln (1/\lambda ))$.
The long-distance behaviour of $W^{in}_{\Lambda}(r) =
O(e^{-Mr})$, with its correlation length $\xi = M^{-1}$ is a
shortlived relic of the initial thermal conditions.  After a time $t_r
= O(M^{-1})\leq t_{sp}$, it is rapidly
supplanted by the behaviour of the expanding long wavelength modes 
$W_{|{\bf k}|<M}(r;t)/W_{|{\bf k}|<M}(0;t)\approx\exp\{-r^{2}/\xi^{2}(t)\}$ where
$\xi^{2}(t) = 4t/M\approx 4/k_{0}^{2}$.
With  $W_{|{\bf k}|<M}(0;t_{sp}) = O(\lambda^{-1})$ non-perturbatively
large and
$W^{in}_{\Lambda}(0)$ and $ W_{\Lambda >|{\bf k}|>M}(0;t_{sp})$ of
order $\lambda^{-1/2}$,
\begin{equation}
n_{M}(t_{sp})\approx  
\frac{1}{\pi\xi^{2}(t_{sp})}[1 + O(\lambda^{1/2}\ln (1/\lambda ) )]
\label{niMf}
\end{equation}
to give Eq.\ref{ntsp}, as promised. 
The error term in the brackets is, in large part, a measure of the fluctuation
vortices that we mentioned earlier.
To check the stability of this density to changes in the
coarse-graining scale $\Lambda = O(M)$ a simple calculation shows that
%\begin{equation}
$\Theta_{M}(t_{sp}) = O(\lambda^{1/2}\ln
(1/\lambda ))\ll 1$
%\label{nstab}
%\end{equation}  
 for weak coupling.  

There are other, less direct, ways to understand why the strings only become
well-defined once $W_{M}(0;t)$ is nonperturbatively large, even
though the density is
independent of its {\it magnitude}. Instead
of a field basis, we can work in a particle basis and measure the
particle production as the transition proceeds.  
Whether we expand with respect to the original
Fock vacuum or with respect to the adiabatic vacuum state, the
presence of a non-perturbatively large peak in $k^2 P(k;t)$ signals
a non-perturbatively large occupation number $N_{k_{0}}\propto 1/\lambda$
of particles at the same wavenumber $k_0$\cite{boyanovsky}. 
With $n_{M}$ of (\ref{niMf}) of order $k_{0}^{2}$ 
this shows that the long wavelength modes can now begin to be treated classically.  

>From a slightly different viewpoint, the Wigner functional only
peaks about the classical phase-space trajectory once the power is
non-perturbatively large\cite{guth,muller} from time $t_{sp}$ onwards.  
More crudely, the diagonal density matrix
elements (field probabilities) are only then significantly non-zero for
non-perturbatively large field configurations
$\phi\propto\lambda^{-1/2}$ like vortices.

So far we have done no more than estimate the density of
coarse-grained line zeroes for a free roll at the time $t_{sp}$ at which
 nothing has frozen in.  
For this {\it global} $O(2)$ theory the damping of domain growth occurs
by the self-interaction effectively  forcing the
negative $m^{2}(t)$ to
vanish so as to produce Goldstone particles.  It is instructive to adopt
the simplest assumption
that will lead to damping, in which we extend the time dependence of the
free-roll mass  to
$m^2(t) = 0$ if $t > t_{sp} $.
The non-perturbatively large $W_{{\bf k}<M}(r;t)$ then has a 
 power spectrum that is the old one, frozen at time $t_{sp}$, upon
which oscillatory behaviour is superimposed.
%$W(r;t)$ undergoes {\it damped} oscillations about its non-perturbatively
%large value at $t=t_{sp}$, which die away on
%a time scale
%$\delta t = O((t_{sp}/M)^{1/2})$. 
%Of course, since $m^{2}(t)\approx0$ does not set in immediately this is 
%a serious underestimate.  
%However, 
There is  no qualitative change
in the vortex density once the damping has taken place.

To improve upon the free-roll result more honestly, 
but retain the Gaussian approximation, the best we can do is a mean-field
approximation or, better, a large-$N$ expansion (for $O(N), N=2$). 
 The main effect of the Goldstone
production is to change the power in the field for very soft
($k\rightarrow 0$) modes, which has no significant affect on 
the position of the peak in
the power spectrum, and hence the density\cite{boyanovsky,LA}.

Whatever, but for varying coefficients of $\lambda$, 
the modes $\chi^{\pm}_{k}$ now satisfy the equation\cite{boyanovsky}
\begin{equation}
\Biggl [ \frac{d^2}{dt^2} + \omega^{2}(t) +  
\lambda\int d \! \! \! / ^3 p 
\, C(p)
[\chi^{+}_{p}(t)\chi^{-}_{p}(t)-1]
\Biggr ]\chi^{\pm}_{k}(t) =0.
\label{modeh2}
\end{equation}
where $\omega^{2}(t) ={\bf k}^2 + m^2(t)$.
A rough estimate of the effects of the interactions prior to
$t_{sp}$ can be obtained
by  retaining only the unstable modes
in the integral for which  the  equations can be written
\begin{equation}
\Biggl [ \frac{d^2}{dt^2} + {\bf k}^2 -\mu^2(t) \Biggr ]\chi_{k}(t)  =0.
\label{modemu}
\end{equation}
%Since the mode with wavenumber $k >0$ stops growing at time $t^{+}_k
%<t_{sp}$, 
%$\mu^{2}(t^{+}_{k}) = {\bf k}^{2}$,
%the free-roll density at $t_{sp}$ must be an overestimate.  To see that
%it is not a large overestimate 
%we
%suppose that, in the vicinity of $t^{+}_k$,
%$\Omega_{k}^{2}(t) = \mu^{2}(t) - {\bf k}^{2}\sim C_{k}^{2}M^{2}[M(t^{+}_k -t)]^{n}$
%for some $n$. 
%Let $S_{k}(t) = \int_{t}^{t^{+}_k }dt'\,\Omega_{k}(t')$.
%As $t$ decreases from $t^{+}_k$
% the relevant solution is, for $m = 1/(n+2)$
%\begin{equation}
%\chi_{k}(t) = B_{k}\bigg(\frac{M S_{k}(t)}{\Omega_{k}(t
%)}\bigg)^{1/2}
%K_{m}(S_{k}(t)) \, e^{S_{k}(t)}
%\label{chiKf}
%\end{equation}
%when $M(t^{+}_k -t)$  is large, with $B_{k} = O(1)$.
%This suggests 
We adopt the approximate hybrid self-consistent equation for $\mu^{2}(t)$, 
\begin{equation}
\mu^{2}(t)\simeq M^{2} - C\lambda \frac{T_{0}M}{(Mt_{sp})^{3/2}}\exp
\bigg(2\int_{0}^{t}dt'\,\mu (t')\bigg),
\label{muh2}
\end{equation}
($C = O(1)$), which has the exponential growth of the WKB solution, but 
non-singular behaviour when $\mu (t)\approx 0$.
The exact  solution to Eq.\ref{muh2} for $t<t_{sp}$ is
$\mu (t) = M\tanh M(t_{sp}-t)$.
%which corresponds to taking $n=1$ and $C^{2}_{k} = 2|k|/M$.
The condition $\mu (t_{sp}) = 0$ reproduces 
$Mt_{sp}= O[\ln (1/\lambda )]$.
The first effect  of the
back-reaction is only to give a time-delay of $O(M^{-1})$,  reducing the
density by an unimportant $O(1/M^{2})$.  

For $t>t_{sp}$ oscillatory modes take over the
correlation function to give oscillations in $W_{\Lambda}(r;t)$ and
its normalised derivatives and, through Eq.\ref{nlambda}, 
to oscillations in the string density. 
Interpolating the mode equation to larger $t$ leads to oscillations with period
$O(M^{-1})$. Explicit dissipative terms are not
necessary to enforce damping since the $O((\langle \phi_{a}({\bf
x})^{2}\rangle )^{2})$ contribution to the mean-field Hamiltonian  acts
like a cosmological constant\cite{boyanovsky}.
In practice
the backreaction rapidly forces $\mu^{2}(t)$ towards zero if the
coupling is not too small\cite{boyanovsky}.
Since the effect of imposing oscillations upon the frozen power
spectrum is to leave the density, independent of normalisation,
qualitatively unchanged in the nature of its $\lambda$ dependence
it makes sense to identify the density at $t_{sp}$ with the
density at freeze-in. 

All the results above were for the instantaneous quench. We now
return to the original problem of slower quenches,
in which the defects begin to  freeze in by
time $t<t_Q$.
For a {\it
free} roll,  with $m^{2}(t)$ as given before,
the mode equation Eq.\ref{mode1} is  now of the form 
\begin{equation}
\Biggl [ \frac{d^2}{dt^2} + {\bf k}^2  -\frac{M^{2}t}{t_Q} \Biggr ]\chi^{\pm}_{k}(t)  =0,
\label{mode1s}
\end{equation}
 The exponentially growing modes that appear when
$t>t^{-}_{k} = t_{Q}k^{2}/M^{2}$ lead to the approximate solution
\begin{equation}
W(r;t)\propto
\frac{T}{M|m(t)|}\bigg(\frac{M}{\sqrt{tt_Q}}\bigg)^{3/2}e^{\frac{4Mt^{3/2}}
{3\sqrt{t_Q}}}
e^{-r^{2}/\xi^{2}(t)}
\label{Wexp}
\end{equation}
where 
$\xi^{2}(t) = 2\sqrt{tt_Q}/M$.
The provisional freeze-in time $t_{sp}$ is then, effectively,
the time it takes to reach the transient groundstate $\phi_{0}^{2}(t)= -m^{2}(t)/\lambda$.
That is, $W(0;t_{sp}) = O(\phi_{0}^{2}(t_{sp}))$, giving, for 
$Mt_{Q} < (1/\lambda)$,
\begin{equation}
Mt_{sp} \simeq (Mt_{Q})^{1/3}(\ln (1/\lambda ))^{2/3}
\simeq Mt_{C}(\ln (1/\lambda ))^{2/3},
\label{tfs}
\end{equation}
greater than $Mt_{C}$.

At this qualitative level the correlation length at the spinodal time is
\begin{equation}
M^{2}\xi^{2}(t_{sp})\simeq (Mt_{Q})^{2/3}(\ln (1/\lambda ))^{1/3}.
\label{chiss2}
\end{equation}
The effect of the other modes is larger than for the instantaneous
quench,  giving, at $t=t_{sp}$
\begin{equation}
n_{M}
=  
\frac{1}{\pi\xi^{2}(t_{sp})}[1 +
O(\lambda^{1/2}(Mt_{Q})^{4/3}(\ln (1/\lambda))^{-1/3})].
\label{nisMf}
\end{equation}
The error term is due to fluctuation vortices.  It is small if $Mt_Q$ is
 a very small power of $\lambda^{-1}$. However, it  is
more comfortable to take $Mt_Q = O((\ln (1/\lambda ))^{\alpha})$
($\alpha > 1$) or, equivalently, as a power (greater than one) of
$Mt_{sp}$, still permitting $Mt_Q >Mt_{sp}$. 
Then  for small enough $\lambda$, we recover
$n_{M}$ of Eq.\ref{enslow}.
As could be anticipated from Eq.\ref{nisMf},
%\begin{equation}
$\Theta_{M}(t_{sp})\approx 
O(\lambda^{1/2}(Mt_{Q})^{4/3}(\ln (1/\lambda ))^{-1/3})$,
%\label{nstab4}
%\end{equation}  
%As a positive power of $\lambda$ times logarithms, this will be 
small enough to give an identifiable network on freezing in
provided $Mt_Q$ is constrained as above.
Equivalently, by time $t_{sp}$ we have nonperturbatively large
occupation numbers for the long wavelength modes, as evidenced by
the WKB solutions.

The density of  Eq.\ref{tom} is {\it larger} than that of
Eq.\ref{nisMf} by the factor $(\ln (1/\lambda ))^{1/3}$ which 
 is, in principle, a large number even if in
practice it is only a small multiple.
  However, Eq.\ref{tom} is a misleading upper bound when considering definition of the
network since a simple calculations shows that, at time $t_C$,
$\Theta_{M}(t_{C}) = O(1)$. 
At this time, with no large
occupation numbers, a classical description of strings is inappropriate.  
The bound set by causality is not saturated in our weak-coupling
theory.

For $t<t_{sp}$ the backreaction is more complicated than for the
instantaneous quench.  A rough approximation is obtained by 
iterating the free-field behaviour.
The mode $\chi_{k}(t)$ then grows until time $t^{+}_k$, defined by
\begin{equation}
k^2 = \frac{M^2 t^{+}_k}{t_Q}
-\frac{C \lambda T}{M^2}\bigg(\frac{M}{\sqrt{t^{+}_k t_Q}}\bigg)^{3/2}
e^{4M(t^{+}_k )^{3/2}/3\sqrt{t_Q}}
\label{Wexp3}
\end{equation}
The spinodal time $t_{sp}$ is the solution to
Eq.\ref{Wexp3} for $k = 0$.  In terms of $t_{sp}$ the mode $\chi_{k}(t)$
stops growing at a time $\Delta t(k)$ before it would have stopped
had the backreaction been  instantaneous, where
$M\Delta t(k) = O((k^2 /M^2)(t_Q/t_{sp})^{3/2})$.
   From Eq.\ref{Wexp} this can be reinterpreted as a
timelag $\Delta t$, where
$M\Delta t = (t_Q/t_{sp})^{1/2}$.
Although this is logarithmically large, relatively to $t_{sp}$ it is
the same as for the instantaneous quench.
We have to treat these ratios with care, since there is no
self-consistency in them, but they continue to suggest that the
initial effect of the back-reaction is benign.  

It seems that the
initial effect of the backreaction {\it after} time $t_{sp}$ is to set up
oscillations of period $O(M^{-1}t_Q/t_{sp})$ followed,
once $\mu^{2}(t)\approx 0$, by further damped oscillations over a time scale 
$\delta t = O((\sqrt{t_Q t_{sp}}/M)^{1/2})$.
As for the instantaneous quench, the density at $t_{sp}$ determines 
the density at freeze-in.

The uncertainty in string length because of the
fluctuations in the short wavelength modes persists at the same
level, increasing as the quench rate is slowed.
Once the rate is slow enough the WKB
approximation manifestly breaks down. This is almost certainly the
case if the growing modes catch up with the moving minima within the
Ginzburg regime by time
$t_G = O(\lambda t_Q )$.
%If this behaviour is
%inserted into Eq.\ref{nstab4} then we see 
Simple inspection shows that fluctuation vortices
dominate the density and
$\Theta_{M}(t_{G})$ is large.  With no strong peaking in the power
$k^{2}P(k)$ and no large occupation numbers that betoken classical
behaviour the strings do not provide a well-defined network.  This
is despite their low density, which can be made compatible with the equilibrium
fluctuation density $n_{G} = O(\lambda A^{-1})$ that was originally
considered to be the relevant density for defect
formation\cite{kibble1,ray}.

We thank Tim Evans, Dan
Boyanovsky, Rich Holman, Hector de Vega and Woytiech Zurek for 
fruitful discussions.
G.K. would like to thank the Greek State Scholarship Foundation (I.K.Y.) for
financial support. 
This work was supported in part by the European Commission under the
Human Capital and Mobility programme, contract number ERB-CHRX-CT94-0423.

\end{document}